\documentclass[useAMS,usenatbib]{mn2e}

 \usepackage{times}

\usepackage{amsmath}
\usepackage{amssymb}
\usepackage{color}
\usepackage{graphicx}
\usepackage{mathtools, cuted}
\definecolor{darkgreen}{rgb}{0.33, 0.42, 0.18}\newcommand{\C}{\mathbf{C}}
\newcommand{\Cinv}{\mathbf{C}^{-1}}

\newcommand{\E}{\mathbf{E}}
\renewcommand{\epsilon}{\varepsilon}

\newcommand{\f}{\mathbf{f}}

\newcommand{\inv}{^{-1}}
\renewcommand{\k}{{\bmath{k}}}

\newcommand{\N}{\mathbf{N}}

\renewcommand{\Re}{\operatorname{Re}}
\newcommand{\revision}[1]{ #1}

\newcommand{\tr}{\operatorname{tr}}

\title[Unbiased contaminant removal for 3D power]{Unbiased contaminant removal for 3D galaxy power spectrum measurements}

\author[B. Kalus, W. J. Percival, D. J. Bacon and L. Samushia]{B. Kalus$^{1}$\thanks{E-mail:
benedict.kalus@port.ac.uk}, W. J. Percival$^{1}$, D. J. Bacon$^{1}$ and L. Samushia$^{2,1,3}$ %\footnotemark[1]\thanks{This file has been amended to
\\
$^{1}$Institute of Cosmology \& Gravitation, Dennis Sciama Building, University of Portsmouth, Portsmouth, PO1 3FX, UK\\
$^{2}$Department of Physics, Kansas State University, 116, Cardwell Hall, Manhattan, KS, 66506, USA\\
$^{3}$National Abastumani Astrophysical Observatory, Ilia State University, 2A Kazbegi Ave., GE-1060 Tbilisi, Georgia}
\begin{document}

\date{Accepted . Received ; in original form \today}

\pagerange{\pageref{firstpage}--\pageref{lastpage}} \pubyear{2016}

\maketitle

\label{firstpage}

\begin{abstract}
	We assess and develop techniques to remove contaminants when calculating the 3D galaxy power spectrum. We separate the process into three separate stages: (i) removing the contaminant signal, (ii) estimating the uncontaminated cosmological power spectrum, (iii) debiasing the resulting estimates. For (i), we show that removing the best-fit contaminant (\revision{\textit{mode subtraction}}), and setting the contaminated components of the covariance to be infinite (\textit{mode deprojection}) are mathematically equivalent. For (ii), performing a Quadratic Maximum Likelihood (QML) estimate after \textit{mode deprojection} gives an optimal unbiased solution, although it requires the manipulation of large $N_{\rm mode}^2$ matrices \revision{($N_{\rm mode}$ being the total number of modes)}, which is unfeasible for recent 3D galaxy surveys. Measuring a binned average of the modes for (ii) as proposed by \citet*[FKP]{Feldman} is faster and simpler, but is sub-optimal and gives rise to a biased solution. We present a method to debias the resulting FKP measurements that does not require any large matrix calculations. We argue that the sub-optimality of the FKP estimator compared with the QML estimator, caused by contaminants is less severe than that commonly ignored due to the survey window.
\end{abstract}


\section{Introduction}

Galaxy surveys provide a rich store of information about the nature of the Universe,
allowing us to constrain cosmological models with baryon acoustic oscillations
(BAO), gravitational models with redshift space distortions (RSD) and
inflationary models with primordial non-Gaussianity.  A basic statistic
containing large-scale structure information is the galaxy power spectrum $P(k)$,
which is the 2-point function of the Fourier transformed density field.  Future
large-scale structure surveys, such as the  Dark Energy Spectroscopic Instrument
survey \citep[DESI]{Schlegel:2011zz,Levi:2013gra}, Euclid
\citep{Laureijs:2011gra}\footnote{www.euclid-ec.org} and the Square Kilometre
Array (SKA) \footnote{www.skatelescope.org}, will probe larger volumes,
therefore allowing us to measure more Fourier modes of the galaxy density field. 

The observed galaxy field can be contaminated with fluctuations of
non-cosmological origin, such as variations due to the galactic extinction and the
stellar density. Often the contaminants are not known exactly (e.g. we may
know the shape of the spurious mode but may not know its exact amplitude) which
makes their exact removal impossible. These modes have the potential to strongly
bias cosmological constraints derived from the clustering measurements, so we need to 
correct or suppress these misleading modes in a responsible way. 

We now introduce the basic mathematical problem that we wish to solve and
introduce the main methods of removing contaminants discussed in literature. We
assume that we have measured the galaxy density field as real numbers in configuration space, which we (fast)
Fourier transform to obtain a Hermitian density field $F(\k)$. Furthermore, we
assume that the contamination can be described by another Hermitian field
$f(\k)$, such that the true density field is given by
\begin{equation}
	D(\k)=F(\k)-\revision{\varepsilon_\mathrm{true}}f(\k),
	\label{eq:Fassum}
\end{equation}
\revision{with $\varepsilon_\mathrm{true}$ unknown.}
In cases with multiple contaminants (which we label with capital Latin indices),
we extend Eq.~\eqref{eq:Fassum} to
\begin{equation}
	D(\k)=F(\k)-\sum_A\revision{\bvarepsilon^\mathrm{(true)}_{A}} f_A(\k).
\end{equation}
Furthermore, we assume that $F(\k)$ and $f(\k)$ are uncorrelated, which is a
valid assumption for most sources of systematics since they originate from our
Galaxy or due to telescope effects. 
Large scale surveys will reduce the current sample variance limitation on the power spectrum on scales where the systematic errors have a significant impact. As a consequence, having control of these systematics is a key requirement to provide accurate cosmological measurements. 

In order to investigate techniques for estimating the power spectrum in the presence of contaminants, we separate the process into three separate stages: (i) removing the contaminant signal, (ii) estimating the uncontaminated cosmological power spectrum, (iii) debiasing the resulting estimates. Two techniques are in common usage for removing the contaminant signal (i): The first is \revision{\textit{mode subtraction}} (cf. Sec.~\ref{sec:systrem}~and Sec.~\ref{sec:debias}), where contaminants are removed by fitting the amplitude of the contaminant field $f(\k)$ to the data and simply subtracted off from $F(\k)$. The second is \textit{mode deprojection} \citep{Rybicki}, which is based on assigning infinitely large covariances to contaminated modes, thus removing them from any analysis. \revision{In our nomenclature, a mode is a linear combination of Fourier modes rather than a single $\k$-mode. This is reflected in the naming of \textit{mode subtraction} and \textit{mode deprojection}. This choice of names shall distinguish the \textit{mode subtraction} technique from a third technique for removing the contaminant signal, called \textit{template subtraction}, where the observed power spectra are corrected using best-fit amplitudes derived via cross-correlations between the data and the templates. \citet*{Elsner} have shown that this method provides a biased estimate of the power and we will not consider it further in this article.} For (ii), the power spectrum P(k) is commonly estimated by the FKP estimator \citep{Feldman}, which is an approximation to the Quadratic Maximum Likelihood (QML) estimator \citep{Tegmark:1997rp}. As well as being optimal, the QML estimator has the advantage of producing unbiased power spectrum estimates. However, when applying this methodology \revision{to data with $N_\mathrm{mode}$ modes}, one has to calculate, for each bin, a $N_\mathrm{mode} \times N_\mathrm{mode}$ matrix, \revision{and then, after binning the data into $N_\mathrm{bin}$ bins,} an overall $N_\mathrm{bin} \times N_\mathrm{bin}$ normalisation matrix, which makes the application of this methodology unfeasible for future surveys with increased number of modes $N_\mathrm{mode}$. In this work, we suggest a modified FKP-style \revision{\textit{mode subtraction}} approach. We show that this technique can be made unbiased and, on a mode-by-mode basis, is mathematically identical to \textit{mode deprojection}. The FKP estimator with debiased \revision{\textit{mode subtraction}} is not optimal in that it discards more information than the full QML estimator, but we expect that, in realistic cases, this loss of information will be small.

The outline of this paper is as follows: We provide an introduction to power
spectrum estimation in Sec.~\ref{sec:notation}, introducing the QML and FKP estimators. We introduce the systematics removal techniques, \textit{mode deprojection} and \revision{\textit{mode subtraction}}, in sections \ref{sec:BMP} and \ref{sec:systrem}, respectively, and we show that before normalisation their resulting power spectra are the same. These are extended to multiple contaminants in Appendices \ref{sec:MPmultderiv} \&\;\ref{sec:TSmultideriv}, respectively. We introduce a new normalisation factor in Sec.~\ref{sec:debias} for a single contaminant and compare it to the normalisation of the quadratic maximum likelihood (QML) estimator of \citet{Tegmark:1997rp}. This derivation is extended to allow for a non-diagonal covariance in Appendix \ref{sec:nondiag}. We show that we can apply our methodology also to multiple contaminants in Sec.~\ref{sec:multisyst} and test the different methods on simulations in Sec.~\ref{sec:eg}. We conclude in Sec.~\ref{sec:conclusion}.

\section{Power spectrum estimators}
\label{sec:notation}
In this section, we review two basic power spectrum estimators: the \textit{quadratic maximum likelihood} (QML) estimator \citep{Tegmark:1997rp} and the simplified FKP estimator \citep{Feldman}, to which QML reduces in the limit of uncorrelated modes with equal noise per mode in each bin. Even without considering any contaminants, the FKP estimator is easier to implement and is used in most recent analyses of large-scale structure, while the QML estimator is optimal but difficult to implement especially on smaller scales.

The \textit{quadratic maximum likelihood} (QML) estimator \citep{Tegmark:1997rp} is given by
\begin{equation}
	\widehat P(k_i)=\sum_{j}\N_{ij}\inv\mathbf{p}_j,
	\label{eq:QuadEst}
\end{equation}
where the power is a convolution of the inverse of a normalisation matrix $\N_{ij}$ and a weighted two-point function  
\begin{equation}
	\mathbf{p}_j\equiv \sum_{\alpha, \beta} F^\ast(\k_\alpha) \E_{\alpha\beta}(k_j) F(\k_\beta).
	\label{eq:pundef}
\end{equation}
The weight is given by the estimator matrix
\begin{equation}
	\E(k_j)=-\frac{\partial \Cinv}{\partial P(k_j)},
\end{equation}
which describes how the inverse of the density field covariance matrix $\C$ changes with respect to the prior of the power spectrum of the respective bin. If the QML normalisation is proportional to the Fisher information, i.e.
\begin{align}
	\N_{ij}&=\tr\left\lbrace \Cinv\frac{\partial\C}{\partial P(k_i)}\Cinv\frac{\partial\C}{\partial P(k_j)}\right\rbrace,
	\label{eq:Ndef}
\end{align}
the QML estimator is the optimal maximum likelihood estimator of the variance of a field that obeys a multivariate Gaussian distribution \citep{Tegmark:1997rp}. Assuming a Gaussian density field $D(\k)$, the QML estimator therefore provides an estimate of the power spectrum with minimal errors.

Under the assumption that all modes are independent, the covariance of the density field is given by the power spectrum (and the Kronecker delta $\delta_{\mu\nu}$):
\begin{equation}
	\C_{\mu\nu}=\delta_{\mu\nu}P(k_{\mu}).
	\label{eq:DiagCov}
\end{equation}
We assume for the derivative of $\C$ with respect to $P(k_i)$ that it is unity if the modes $\k_\alpha$ and $\k_\beta$ are equal and contained in the bin $\Bbbk_i$, and zero otherwise, which we write using the Heaviside function $\Theta$ as:
\begin{equation}
	\frac{\partial\C_{\alpha\beta}}{\partial P(k_i)}=\delta_{\alpha\beta}\Theta(\k_\alpha\in \Bbbk_i)\equiv\delta_{\alpha\beta}\Theta_{\alpha i}.
	\label{eq:Cderiv}
\end{equation}
Given Eq.~\eqref{eq:DiagCov} and \eqref{eq:Cderiv}, we find
\begin{align}
	\E_{\alpha\beta}(k_j)=\frac{\delta_{\alpha\beta}}{P^2(k_\alpha)}\Theta_{\alpha j}
\end{align}
and
\begin{align}
	\N_{ij}=\frac{N_{\Bbbk_i}}{P^2(k_i)}\delta_{ij},
	\label{eq:FKPNorm}
\end{align}
where $N_{\Bbbk_i}$ is the total number of modes in a given bin $\Bbbk_i$. Hence the QML estimator of Eq.~\eqref{eq:QuadEst} reduces to the FKP estimator \citep*{Feldman} under the assumption that the covariance is constant within the $\k$-bin, where several modes (labelled with Greek indices) are combined into bins (denoted by $\Bbbk_i$ and distinguished with lower case Latin indices) and the absolute values squared of the density field of each bin are summed:
\begin{equation}
	\widehat{P}(k_i)=\frac{1}{N_{\Bbbk_i}}\sum_{\k_\alpha\in\Bbbk_i}\left\vert F(\k_\alpha)\right\vert^2.
	\label{eq:FKP}
\end{equation}
The difference here is that the QML estimator uses a prior of the power spectrum $P(k_\alpha)$ to weight contributions from each mode optimally, which means that the covariance of the power spectrum is minimal.
The FKP estimator is commonly applied even when the assumptions of Eq.~\eqref{eq:DiagCov} to \eqref{eq:FKPNorm} are not valid.

\section{Removing Contaminants: mode deprojection}
\label{sec:BMP}
We now describe how \textit{mode deprojection} can be applied to estimate the 3D galaxy power spectrum. The method was first suggested in \citet{Rybicki} in the context of noisy, irregularly sampled data. Applications and extensions to angular power spectra can be found for WMAP data in \citet{Slosar:2004fr}, for SDSS-III data in \citet{Ho:2012vy}, for photometric quasars of the XDQSOz catalogue in \citet{Leistedt:2014wia} and \citet{Leistedt:2014zqa} and for 2D galaxy clustering in general in \citet{Elsner}. We use the notation of \citet{Elsner} for consistency.

Suppose we estimate the power spectrum using QML and that there is only a single contaminant. Then one can suppress contaminated modes in the covariance matrix updating the covariance matrix as \citep{Elsner}
\begin{equation}
	\C_{\alpha\beta}\rightarrow\tilde\C_{\alpha\beta}=\C_{\alpha\beta}+\lim_{\sigma\rightarrow\infty}\sigma f(\k_\alpha)f^\ast(\k_\beta),
	\label{eq:MP}
\end{equation}
i.e. letting the covariances of contaminated modes tend to infinity.
Making use of the Sherman-Morrison matrix inversion lemma \citep{Sherman}, one can see that (if $f(\k)\neq 0\;\forall\k$) the inverse updated covariance matrix converges to
\begin{equation}
	\tilde\C\inv_{\alpha\beta}=\C\inv_{\alpha\beta}-\frac{\sum_{\mu\nu}\C_{\alpha\mu}\inv f(\k_\mu)f^\ast(\k_\nu)\C_{\nu\beta}\inv}{\sum_{\mu\nu} f^\ast(\k_\mu)\C\inv_{\mu\nu}f(\k_\nu)}.
	\label{eq:ShermanMorrisonLimit}
\end{equation}
Now supposing that the modes are independent, i.e. Eq.~\eqref{eq:DiagCov} holds, we can insert it into Eq.~\eqref{eq:ShermanMorrisonLimit} so that
\begin{equation}
	\tilde\C\inv_{\alpha\beta}=\frac{\delta_{\alpha\beta}}{P(k_\alpha)}-\frac{1}{R_P} \frac{f(\k_\alpha)f^\ast(\k_\beta)}{P(k_\alpha)P(k_\beta)}
	\label{eq:Ctildeinv}
\end{equation} where we have defined 
\begin{equation}
	R_P\equiv \sum_\mu\frac{\vert f(\k_\mu)\vert^2}{P(k_\mu)},
\end{equation}
for simplicity.
Taking the derivative of Eq.~\eqref{eq:Ctildeinv} with respect to $P(k_i)$, we obtain the updated estimator matrix\footnote{writing $f_\alpha\equiv f(\k_\alpha)$ and $P_\alpha\equiv P(k_\alpha)$ to save space}
\begin{equation}
	\tilde\E_{\alpha\beta}(k_j)=\frac{\delta_{\alpha\beta}}{P^2_\alpha}\Theta_{\alpha j}
	-\frac{1}{R_P} \frac{f_\alpha f^\ast_\beta}{P_\alpha P_\beta}\left(\frac{\Theta_{\alpha j}}{P_\alpha}+\frac{\Theta_{\beta j}}{P_\beta}-\frac{t_j}{R_P}\right),
	\label{eq:Ealphabetaj}
\end{equation}
where 
\begin{equation}
	t_i\equiv \sum_{\k_\alpha\in\Bbbk_i}\frac{\vert f(\k_\alpha)\vert^2}{P^2(k_\alpha)}.
\end{equation}
After inserting Eq.~\eqref{eq:Ealphabetaj} into Eq.~\eqref{eq:pundef}, we obtain for the two point function
\begin{align}
	\mathbf p_i
	=& \sum_{\k_\alpha\in\Bbbk_i}\left\lbrace\frac{\vert F(\k_\alpha)\vert^2}{P^2(k_\alpha)}-\frac{2}{R_P}\Re\left[S_P \frac{F^\ast(\k_\alpha)f(\k_\alpha)}{P^2(k_\alpha)}\right]\right.\nonumber\\
	&\left.+\frac{\vert S_P\vert^2}{R_P^2}\frac{\vert f(\k_\alpha)\vert^2}{P^2(k_\alpha)}\right\rbrace\nonumber\\
		=& \sum_{\k_\alpha\in\Bbbk_i}\frac{\left\vert F(\k_\alpha)-\frac{S_P}{R_P}f(\k_\alpha)\right\vert^2}{P^2(k_\alpha)},
	\label{eq:punnorm}
\end{align}
where we have defined
\begin{equation}
	S_P\equiv\sum_{\k_\alpha}\frac{F^\ast(\k_\alpha)f(\k_\alpha)}{P_\alpha}.	
\end{equation}
$S_P$ is real, because $F(\k)$ and $f(\k)$ are Hermitian fields with real Fourier transforms. 

Eq.~\eqref{eq:punnorm} is in a considerably simpler form than Eq.~\eqref{eq:pundef} and does not require calculating many matrix elements of the estimator matrix $\E$. We show in the next section that we can consider this equation as a best-fit of the contaminants in the data.

We can normalise the updated mode deprojected QML estimator by replacing $\C$ by $\tilde\C$ in Eq.~\eqref{eq:Ndef}. As the term that suppresses contaminated modes from the covariance matrix in Eq.~\eqref{eq:MP} does not depend on the power $P(k)$, we have $\frac{\partial\tilde\C_{\alpha\beta}}{\partial P(k_i)}=\frac{\partial\C_{\alpha\beta}}{\partial P(k_i)}$ and hence the normalisation is
\begin{align}
	\tilde\N_{ij}=&\sum_{\alpha\mu\nu\rho}\tilde\C_{\alpha\mu}\inv\delta_{\mu\nu}\Theta_{\mu i}\tilde\C\inv_{\nu\rho}\delta_{\rho\alpha}\Theta_{\alpha j}\nonumber\\
	=&\sum_{\alpha\mu}\vert\tilde\C_{\alpha\mu}\inv\vert^2\Theta_{\mu i}\Theta_{\alpha j}\nonumber\\
	=&\sum_{\alpha\mu}\Theta_{\alpha i}\Theta_{\mu j}\left[\frac{\delta_{\alpha\mu}}{P^2(k_\alpha)}\left(1-\frac{2\vert f(\k_\alpha)\vert^2}{R_P P(k_\alpha)}\right)\right.\nonumber\\
	&\left.+\frac{1}{R_P^2}\frac{\vert f(\k_\alpha)f(\k_\mu)\vert^2}{P^2(k_\alpha)P^2(k_\mu)}\right]
	\label{eq:MPNormLong}
\end{align}
where we have used the Hermitian property of $\tilde{\C}\inv$ in the third equality. In the first term in the square brackets, $\k_\alpha$ has to be in both $\Bbbk_i$ and $\Bbbk_j$, hence we can replace one $\Theta$ with $\delta_{ij}$, such that Eq.~\eqref{eq:MPNormLong} can be written as a diagonal matrix with diagonal elements $\tilde n$ and the outer product of a vector with itself:
\begin{align}
	\tilde\N_{ij}
	=&\sum_{\k_\alpha\in\Bbbk_i}\frac{\delta_{i j}}{P^2(k_\alpha)}\left(1-\frac{2\vert f(\k_\alpha)\vert^2}{R_P P(k_\alpha)}\right)+\frac{t_it_j}{R_P^2}\nonumber\\
	\equiv &\tilde n_i\delta_{ij}+\frac{t_it_j}{R_P^2},
	\label{eq:MPNorm}
\end{align}
This means that we can apply the Sherman-Morrison matrix inversion lemma \citep{Sherman}:
\begin{align}
	\tilde\N\inv_{ij}
	&=\frac{\delta_{ij}}{\tilde n_i}-\frac{1}{R_P^2+\sum_\ell \frac{t_\ell^2}{\tilde n_\ell}}\frac{t_i}{\tilde n_i}\frac{t_j}{\tilde n_j}.
	\label{eq:Ndecomp}
\end{align}
As $\tilde\N\inv$ is not diagonal, it does not reduce to a simple FKP style estimator, i.e. if we have $N_\mathrm{bin}$ bins, we have to calculate for each bin the $N_\mathrm{mode}\times N_\mathrm{mode}$ estimator matrix $\E$ and we have to invert the $N_\mathrm{bin}\times N_\mathrm{bin}$ normalisation matrix. This is not feasible for 3D clustering, because of the large number of modes to be considered, especially if we want to choose narrow bins. Including several contaminants makes it even more costly. 

One way around this is a new framework introduced by \citet{Leistedt:2014wia} which they call
\textit{extended mode projection} and that \revision{selectively removes modes based on cross correlations with the data. However, this procedure} reintroduces a small bias \citep{Elsner}. 

Another possibility is using \revision{the methodology of} the SDSS-III Baryon Oscillation Spectroscopic Survey (BOSS)-collaboration\revision{, which is similar to that described in the next section, but applied} at the power spectrum level. However, this method is also biased \citep{Elsner}. Although \citet{Rossinprep}
show that, for the Completed SDSS-III Baryon Oscillation Spectroscopic Survey
(BOSS DR12), the bias is much smaller than the statistical uncertainty, it was shown in the appendix of \citet{Ross:2012qm} that the bias is significant
when one attempts to correct for many systematics. Furthermore, we expect
smaller statistical uncertainties with future surveys, so in the next two
sections we consider a computationally cheaper way of removing this small bias.

\section{Removing Contaminants: \revision{mode} subtraction}
\label{sec:systrem}

Here we will consider \revision{\textit{mode subtraction}} and its link to \textit{mode
deprojection}. In order to remove contaminants we start by \revision{treating the true, but unknown, amplitude of the contamination $\varepsilon_{\rm true}$ in Eq.~\eqref{eq:Fassum} as} a free parameter
$\varepsilon$, so that \revision{an estimate of the true density field $D(\k)$} reads 
\begin{equation}
	\revision{\widehat D}(\k)=F(\k)-\varepsilon f(\k).
	\label{eq:vartemp}
\end{equation}
Note that this is different to the \textit{template subtraction} method introduced by \citet{Ho:2012vy}, which is used by the BOSS collaboration and works entirely at the level of power spectra, whereas Eq.~\eqref{eq:vartemp} works at the map level.
We can write a simplified model of the Gaussian likelihood whose maximum is given by the QML (cf. Eq.~\eqref{eq:QuadEst} and \citealt{Tegmark:1997rp}) in the approximation of a diagonal covariance matrix, with a small contaminant that does not affect the covariance. This is given by
\begin{equation}
	-2\ln\mathcal L=\ln\left(\prod_\k P(k)\right)+\sum_\k\frac{\vert F(\k)-\varepsilon f(\k)\vert^2}{P(k)}.
	\label{eq:chi2}
\end{equation}
We can therefore find $\varepsilon$ by minimising Eq.~\eqref{eq:chi2}, which is equivalent to simultaneously fitting $\varepsilon$ and the model parameters entering the model power spectrum.
The derivative of $\ln\mathcal L$ with respect to $\varepsilon$ reads
\begin{align}
	\frac{\partial\ln\mathcal L}{\partial\varepsilon}&=\sum_\k\frac{\Re\left[F_f(\k)F^\ast(\k)\right]-\varepsilon\vert F_f(\k)\vert^2}{P(k)}.
\end{align}
This expression is equal to zero and the likelihood maximised if 
\begin{equation}
	\varepsilon^\mathrm{(BF)}=\frac{S_P}{R_P}.
\end{equation}

The uncontaminated estimate of the density field is hence given by
\begin{equation}
	\revision{\widehat D}(\k)=F(\k)-\frac{S_P}{R_P}f(\k),
	\label{eq:FsignalBF}
\end{equation}
and we can estimate the power as
\begin{equation}
	\widehat{P}(k_i)=\frac{1}{N_{\Bbbk_i}}\sum_{\k_\alpha}\left\vert F(\k_\alpha)-\frac{S_P}{R_P}f(\k_\alpha)\right\vert^2.
	\label{eq:PTS}
\end{equation}
This is similar to the \textit{mode deprojection} result of
Eq.~\eqref{eq:punnorm} with a bias, missing the inverse noise matrix convolution
of Eq.~\eqref{eq:QuadEst}. \revision{The bias of this estimate comes about because $S_P$ is correlated with the true density field $D(\k)$. This correlation is similar to that created by the internal linear combination (ILC) method \citep[e.g.][]{Bennett:2003ca} for the analysis of cosmic microwave background (CMB) data.} Based on this knowledge, we build an unbiased
FKP-style estimator in the next section.

\section{An Unbiased FKP-Style Estimator}
\label{sec:debias}

We present in this section a simple, although sub-optimal, way to \revision{remove the bias on} the power spectrum estimate \revision{resulting from imperfectly removing systematics using} either Eq.~\eqref{eq:punnorm} or \eqref{eq:PTS}.
A straightforward way to remove the bias consists of calculating the expectation value of the power from each mode analytically, assuming Eq.~\eqref{eq:Fassum}, and divide out the bias. We start with calculating some useful expectations which we need for the final result, summarised in Table~\ref{tab:expectations}. 
\begin{table}
 \centering
 \begin{minipage}{70mm}
  \caption{Expectation values of quantities entering Eq.~\eqref{eq:debiasfactor}.}
  \label{tab:expectations}
  \begin{tabular}{@{}lr@{}}
 \hline
	$\langle F(\k_\alpha) F^\ast(\k_\beta)\rangle$ & $\delta_{\alpha\beta}P(k_\alpha)+\revision{\varepsilon_\mathrm{true}^2} f(\k_\alpha) f^\ast(\k_\beta)$ \\
	$\langle\varepsilon_\mathrm{BF}\rangle$ & $\frac{\langle S_P\rangle}{R_P}=\revision{\varepsilon_\mathrm{true}}$\\
	$\langle\varepsilon^2_\mathrm{BF}\rangle$ & $\frac{1}{R_P}+\revision{\varepsilon_\mathrm{true}^2}$\\
	$\langle S_P F(\k_\alpha)\rangle$ & $f(\k_\alpha)+R_P f(\k_\alpha)\revision{\varepsilon_\mathrm{true}^2}$\\
\hline
\end{tabular}
\end{minipage}
\end{table}
With these equations at hand, we can calculate the expectation of Eq.~\eqref{eq:punnorm} and \eqref{eq:PTS}, i.e. the two-point function of Eq.~\eqref{eq:Fassum}:
\begin{align}
	&\langle\vert F(\k_\alpha)-\frac{S_P}{R_P} f(\k_j)\vert^2\rangle\nonumber\\
	&=\langle\vert F(\k_\alpha)\vert^2\rangle-\frac{2}{R_P}\langle S_P F(\k_\alpha)\rangle f^\ast(\k_\alpha)+\revision{\langle\varepsilon^2_\mathrm{BF}\rangle}\vert f(\k_\alpha)\vert^2\nonumber\\	
%	&=P(k_\alpha)+\vert f(\k_\alpha)\vert^2-2\frac{\vert f(\k_\alpha)\vert^2}{R_P}-2\vert f(\k_\alpha)\vert^2+\frac{\vert f(\k_\alpha)\vert^2}{R_P}+\vert f(\k_\alpha)\vert^2\nonumber\\
	&=P(k_\alpha)-\frac{\vert f(\k_\alpha)\vert^2}{R_P},
\end{align}
hence, we can build an unbiased estimator of the power by dividing each mode in Eq.~\eqref{eq:punnorm} and \eqref{eq:PTS} by
\begin{equation}
	1-\frac{1}{R_P}\frac{\vert f(\k_\alpha)\vert^2}{P(k_\alpha)}.
	\label{eq:debiasfactor}
\end{equation}
If we want to debias the two-point function using this factor, we have to assume
a prior power spectrum. Note that the QML approach also requires the prior knowledge
of the power spectrum. We will see in Sec.~\ref{sec:eg} that the impact of
adopting a slightly wrong prior is indeed small. Our final estimator of the
power spectrum is then 
\begin{equation}
	\widehat{P}(k_i)=\frac{1}{N_{\Bbbk_i}}\sum_{\k_\alpha}\frac{\left\vert F(\k_\alpha)-\frac{S_P}{R_P}f(\k_\alpha)\right\vert^2}{1-\frac{1}{R_P}\frac{\vert f(\k_\alpha)\vert^2}{P(k_\alpha)}}.
	\label{eq:ubPTS}
\end{equation}
Eq.~\eqref{eq:ubPTS} is one of the key results of this article: this is an extension of the FKP estimator that removes potential contaminants from the data in an unbiased way, without the need for large matrices. Moreover, as it is in the same form as the well established FKP estimator, this can easily be folded into estimators for redshift-space clustering such as those by \citet{Bianchi:2015oia} and \citet{Scoccimarro:2015bla}.

The same debiasing factor can also be derived from the QML Fisher information matrix $\N$, which in the QML approach performs both the debiasing and optimisation effects.  Without binning, Eq.~\eqref{eq:MPNorm} simplifies to
\begin{align}
	\tilde\N_{\alpha\beta}
	=&\frac{\delta_{\alpha \beta}}{P^2(k_\alpha)}\left(1-\frac{2\vert f(\k_\alpha)\vert^2}{R_P P(k_\alpha)}\right)+\frac{1}{R_P^2}\frac{\vert f(\k_\alpha)\vert^2}{P^2(k_\alpha)}\frac{\vert f(\k_\beta)\vert^2}{P^2(k_\beta)}.
	\label{eq:Nnobins}
\end{align}
The difference between the two approaches is that QML provides an unbiased optimal power estimate, whereas Eq.~\eqref{eq:ubPTS} has been constructed such that it is only unbiased, i.e. the powers in the denominators of Eq.~\eqref{eq:Nnobins} act as optimal weights to each mode. If we allow for some information loss within bins, by assuming the expected power is constant within each bin, we can replace $P^2(k_\beta)$ by $P(k_\alpha)P(k_\beta)$, such that
\begin{align}
	\tilde\N_{\alpha\beta}
	=&\frac{\delta_{\alpha \beta}}{P^2(k_\alpha)}\left(1-\frac{2\vert f(\k_\alpha)\vert^2}{R_P P(k_\alpha)}\right)+\frac{1}{R_P^2}\frac{\vert f(\k_\alpha)\vert^2}{P^3(k_\alpha)}\frac{\vert f(\k_\beta)\vert^2}{P(k_\beta)}.
\end{align}
This normalisation is proportional to the Fisher information matrix \citep{Tegmark:1997rp}, from which we marginalise out contributions from other modes by summing over all modes $\k_\beta$:
\begin{align}
	\sum_\beta\tilde\N_{\alpha\beta}
	=&\frac{1}{P^2(k_\alpha)}\left(1-\frac{2\vert f(\k_\alpha)\vert^2}{R_P P(k_\alpha)}\right)+\frac{1}{R_P}\frac{\vert f(\k_\alpha)\vert^2}{P^3(k_\alpha)}\nonumber\\
	=&\frac{1}{P^2(k_\alpha)}\left(1-\frac{\vert f(\k_\alpha)\vert^2}{R_P P(k_\alpha)}\right).
\end{align}
This is exactly Eq.~\eqref{eq:debiasfactor} with a factor of
$\frac{1}{P^2(k_\alpha)}$ that cancels out the difference between
Eq.~\eqref{eq:punnorm} and Eq.~\eqref{eq:PTS}. We have therefore shown that
Eq.~\eqref{eq:ubPTS} is a non-optimal, but unbiased, approximation to using the
QML normalisation with \textit{mode deprojection}. In the limit of narrow bins,
when the power spectrum does not change significantly within the bin,
Eq.~\eqref{eq:ubPTS} is mathematically identical to the QML result. We shall
study the impact of this sub-optimality in examples later in Sec.~\ref{sec:eg}.
In fact, we will argue later that this is actually a weaker effect than many
common approximations applied when using the FKP estimator, such as ignoring
large-scale window effects in the QML approach, when averaging large scale
modes.

Note that\revision{, in the absence of systematics,} we have assumed a diagonal covariance matrix in the derivation of both the \revision{\textit{mode subtraction}} and the debiasing step. In practic\revision{e t}he covariance matrix has off-diagonal terms due to the effect of the survey window. However, this is usually not included when calculating the data power spectrum but, instead, it is included as a convolution in the model power spectrum. We show in Appendix \ref{sec:nondiag} that Eq.~\eqref{eq:ubPTS} still holds in the general case of having a non-diagonal covariance matrix, as long as $R_P$ is generalised as in Eq.~\eqref{eq:RPdef}. \revision{This generalised $R_P$ requires the inversion of the full $N_{\rm mode}^2$ covariance matrix. However, we show in Appendix \ref{sec:diagCapprox} that the effect of assuming a diagonal covariance matrix is either small, or can be corrected for using the covariance matrix, without inversion.} 

\section{Removing multiple Contaminants}
\label{sec:multisyst}
We have shown the equivalence between \textit{mode deprojection} and \textit{debiased \revision{mode} subtraction} for one contaminant. A realistic survey has several sources of potential contaminants, so we show here this equivalence holds for an arbitrary number of templates.
For \textit{mode deprojection}, we have to update the covariance matrix with a sum over all templates, and thus we have to replace Eq.~\eqref{eq:MP} with
\begin{equation}
	\tilde \C_{\alpha\beta}=\C_{\alpha\beta}+\lim_{\sigma\rightarrow\infty}\sigma\sum_{A=1}^{N_\mathrm{sys}}f_A(\k_\alpha)f^\ast_A(\k_\beta).
	\label{eq:MPmult}
\end{equation}
Starting from Eq.~\eqref{eq:MPmult}, we derive in Appendix \ref{sec:MPmultderiv} the unbinned \textit{mode deprojection} power spectrum 
\begin{equation}
	\widehat P(k_\alpha)=\left\vert F(\k_\alpha)-\sum_{AB}\mathbf{S}_A\mathbf{R}\inv_{AB}f_B(\k_\alpha)\right\vert^2,
	\label{eq:multiBMPpower}
\end{equation}
where $\mathbf{R}_{AB}\equiv\sum_{\mu}\frac{f_A^\ast(\k_\mu)f_B(k_\mu)}{P(k_\mu)}$ and $\mathbf{S}_A\equiv \sum_\alpha\frac{f_A(\k_\alpha)F^\ast(\k_\alpha)}{P(k_\alpha)}$ are matrix and vector equivalents of $R_P$ and $S_P$, respectively, in contaminant space.

To apply multiple \revision{mode} subtraction, we extend the likelihood given in Eq.~\eqref{eq:chi2} to
\begin{equation}
	-2\ln\mathcal{L}=\sum_\alpha\frac{\vert F(\k_\alpha)-\sum_A \mathbf{\varepsilon}_A f_A(\k_\alpha)\vert^2}{P(k_\alpha)}.
\end{equation}
Writing $\bvarepsilon$ as a vector, the joint maximum likelihood solution fitting all contaminants is given by (cf. Appendix~\ref{sec:TSmultideriv})
\begin{equation}
	\bvarepsilon^\mathrm{(BF)}=\mathbf{R}\inv\mathbf{S}.
\end{equation}
Note that this would require fitting the amplitude of all contaminants simultaneously.
The absolute value squared of the best fitting signal is hence equal to Eq.~\eqref{eq:multiBMPpower}. Hence, we also do not need large $N_\mathrm{mode}\times N_\mathrm{mode}$ matrices when we have to remove several potential contaminants.

We can calculate the debiasing factor
\begin{align}
	\sum_{j}\tilde\N_{ij}P^2(k_i)=1-\sum_{AB}\frac{f_A(\k_i)\mathbf{R}\inv_{AB}f_B^\ast(\k_i)}{P(k_i)}
\end{align}
analogously to Sec.~\ref{sec:systrem} from the \textit{mode deprojection} normalisation matrix without binning.
 
\section{Testing Contaminant Removal}
\label{sec:eg}
In this section we show how simple contaminants can be removed in power spectrum measurements from simulated density fields, using the hitherto described methodologies. 

\subsection{Gaussian Spike Contaminant}
\label{sec:GaussSpike}

\begin{figure}
\centering
\includegraphics[width=\linewidth]{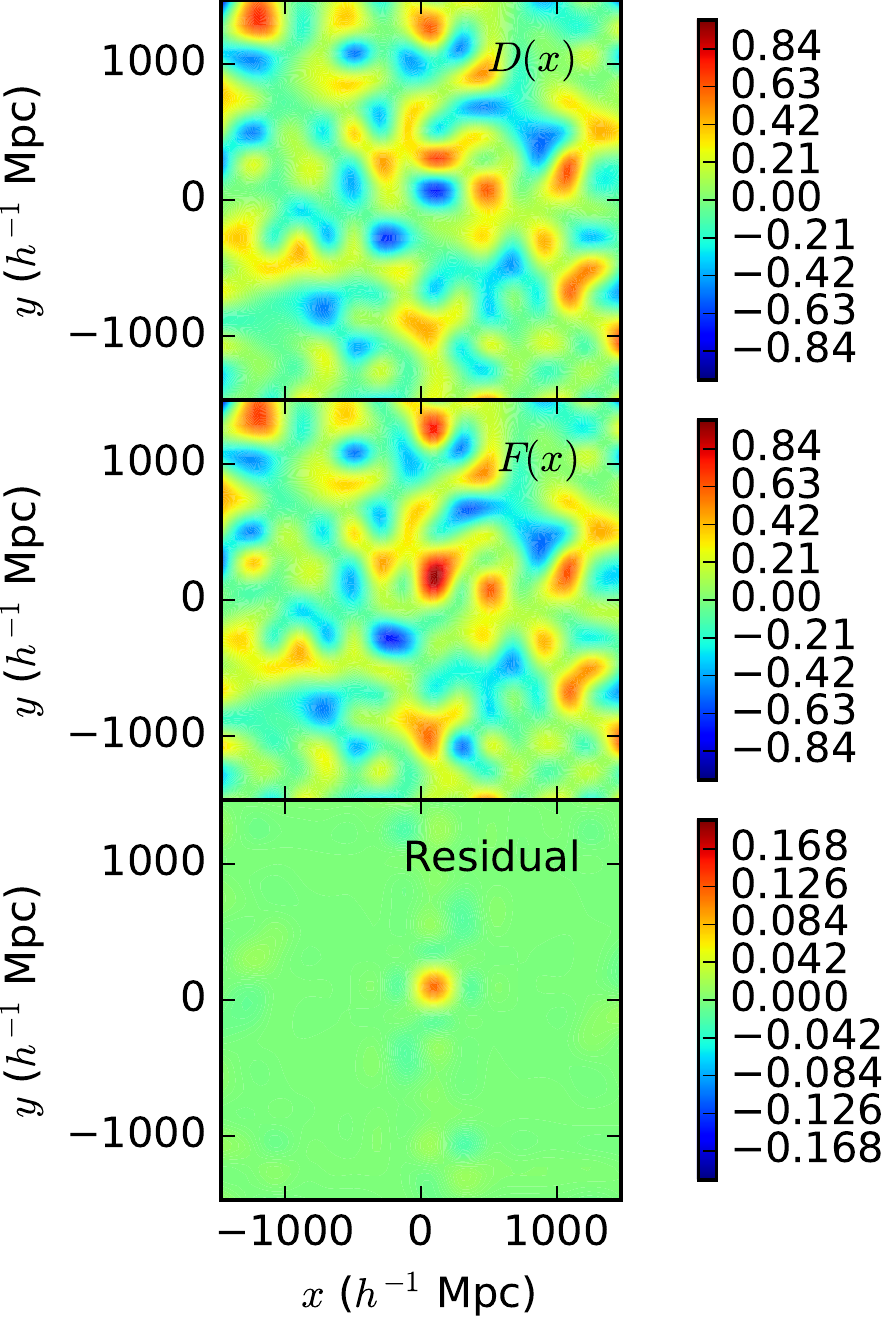}
\caption{A slice through a realisation of a Gaussian random field contaminated with a Gaussian spike used in Sec.~\ref{sec:GaussSpike}. The top panel shows the ``clean" Gaussian random field (corresponding to $D(\k)$ through Fourier transform) in configuration space. In the central panel, we have plotted the contaminated field (Fourier pair of $F(\k)$) with an obvious Gaussian overdensity in the centre. The bottom panel shows \revision{the residual, i.e. the difference of the field after \textit{mode subtraction} (i.e. the Fourier transform of $F(\k)-\epsilon^\mathrm{(BF)} f(\k)$, cf. Eq.~\eqref{eq:PTS}) and the input field}. The best-fitting $\epsilon^\mathrm{(BF)}$ for this particular realisation amounts to 1.078. Although differences between the top and bottom panels are hard to spot by eye, in Fourier space the differences correspond to the bias in the \revision{\textit{mode subtraction}} estimator.}
\label{fig:spikefield}
\end{figure}

As a first test, we generate 3-dimensional Gaussian random fields according to an input power spectrum that we calculate using CAMB \citep{Lewis:1999bs}. Each of these fields consists of a $16\times 16\times 16$ grid, in a box of length $3136h^{-1}\;\mathrm{Mpc}$. An example of such a field is shown in the left panel of Fig.~\ref{fig:spikefield}. We contaminate these Gaussian random fields by adding a real Gaussian spike in $k$-space with width $\sigma^2=10^{-5}h\;\mathrm{Mpc}$, centred around $k=0.01h\;\mathrm{Mpc}^{-1}$, such that its maximum lies within a bin with sufficiently good statistics. The Fourier transform of this contaminant field is again a Gaussian spike in the centre of the box with some long wavelength fluctuations around it. The amplitude of the real part over-density in $\k$-space is 100, thus having the same order of magnitude as the ``true" density field. An example of this setup can be seen in the central panel of Fig.~\ref{fig:spikefield}. We calculate four different power spectra:
\begin{enumerate}
	\item We do not account for the contaminants and just average the absolute values squared of the density field in each bin (cf. Eq.~\eqref{eq:FKP}).
	\item We perform a na\"ive \revision{\textit{mode subtraction}}, i.e. we subtract off the template, but do not debias the two-point function (cf. Eq.~\eqref{eq:PTS}). 
	\item We debias the previous power spectrum by applying Eq.~\eqref{eq:debiasfactor}.
	\item We use the full QML estimator with \textit{mode deprojection}.
\end{enumerate}
In the cases (ii) to (iv), we have to assume a prior power spectrum, which we take as equal to the input power. We shall test the effect of this assumption with the next example. As each bin contains modes with a range of different $\k$-values, we have to clarify what we mean by the prior power spectrum $P(k_i)$ for a specific bin. We find that the power spectrum measurements are closest to the input values, when we assume that the input power spectrum $P(k_i)$ is given by the average of the prior power spectrum values for each mode in the respective bin, i.e.
\begin{equation}
	P(k_i)\equiv \frac{1}{N_{\Bbbk_i}}\sum_{\k_\alpha\in\Bbbk_i}P(k_\alpha).
\end{equation}

\begin{figure}
\centering
\includegraphics[width=\linewidth]{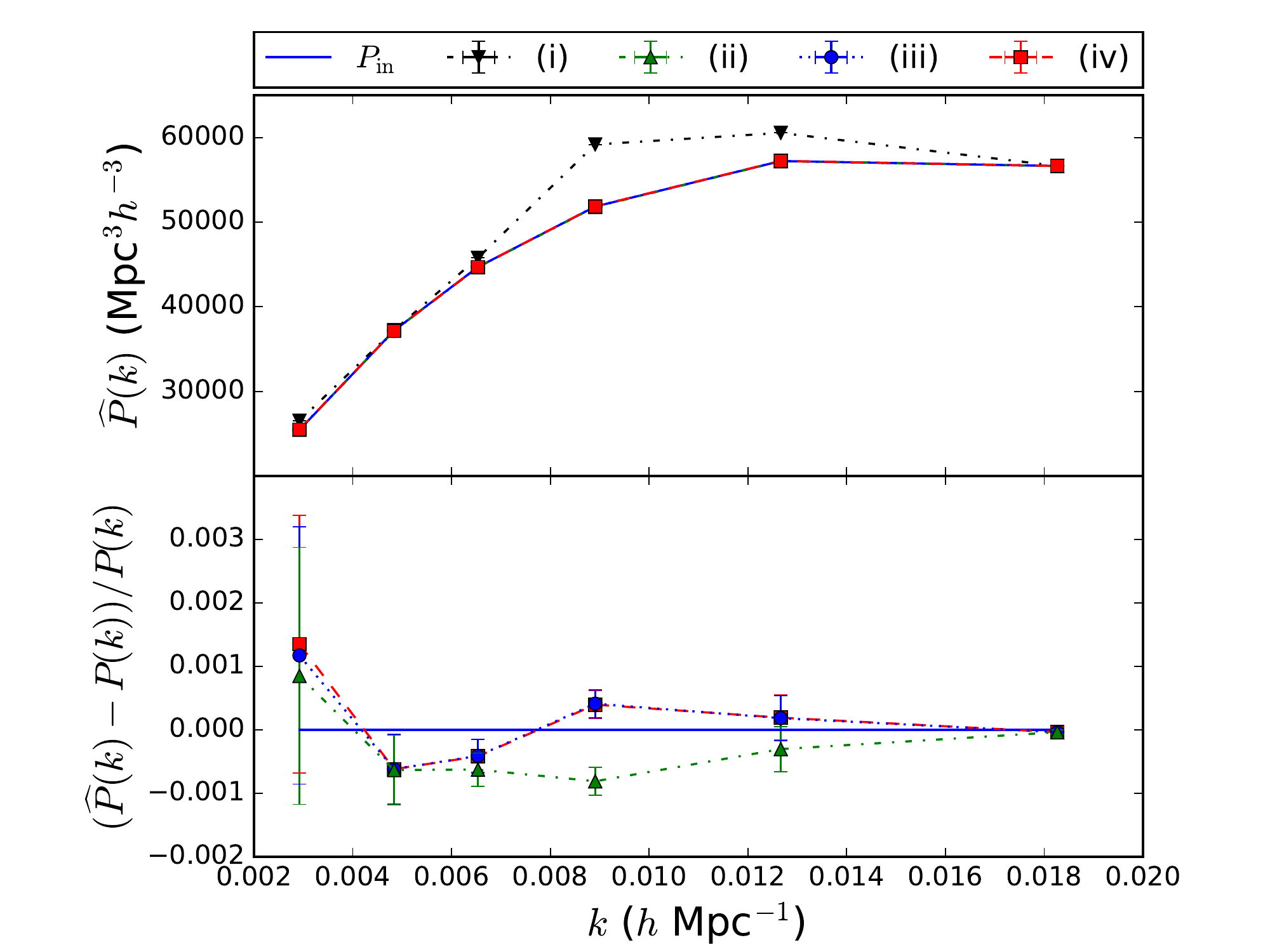}
\caption{Means and standard deviations of the power spectra of 70,000 realisations of Gaussian random fields contaminated with a real Gaussian spike. The top panel shows the input power spectrum as a solid blue line, as well as the power spectra obtained with methods (i)-(iv) as described in Sec.~\ref{sec:GaussSpike}. In the lower panel, we plot fractional errors for methods (ii)-(iv). }
\label{fig:spike}
\end{figure}

In Fig.~\ref{fig:spike}, we can clearly see an increase of power in the bins around $k=0.01$ in case (i). Subtracting off the template in the na\"ive way (method (ii)) is biased in the bins affected by the spike. However, this bias is only a 1 part in a thousand effect. Methods (iii) and (iv) both reproduce the input power spectrum well, removing the bias. A significant difference between their error bars cannot be observed. It is therefore sufficient in this case to use the FKP-style estimator we introduced in Sec.~\ref{sec:debias}.

\subsection{Single Contaminated Mode}
\label{sec:scm}

\begin{figure}
\centering
\includegraphics[width=\linewidth]{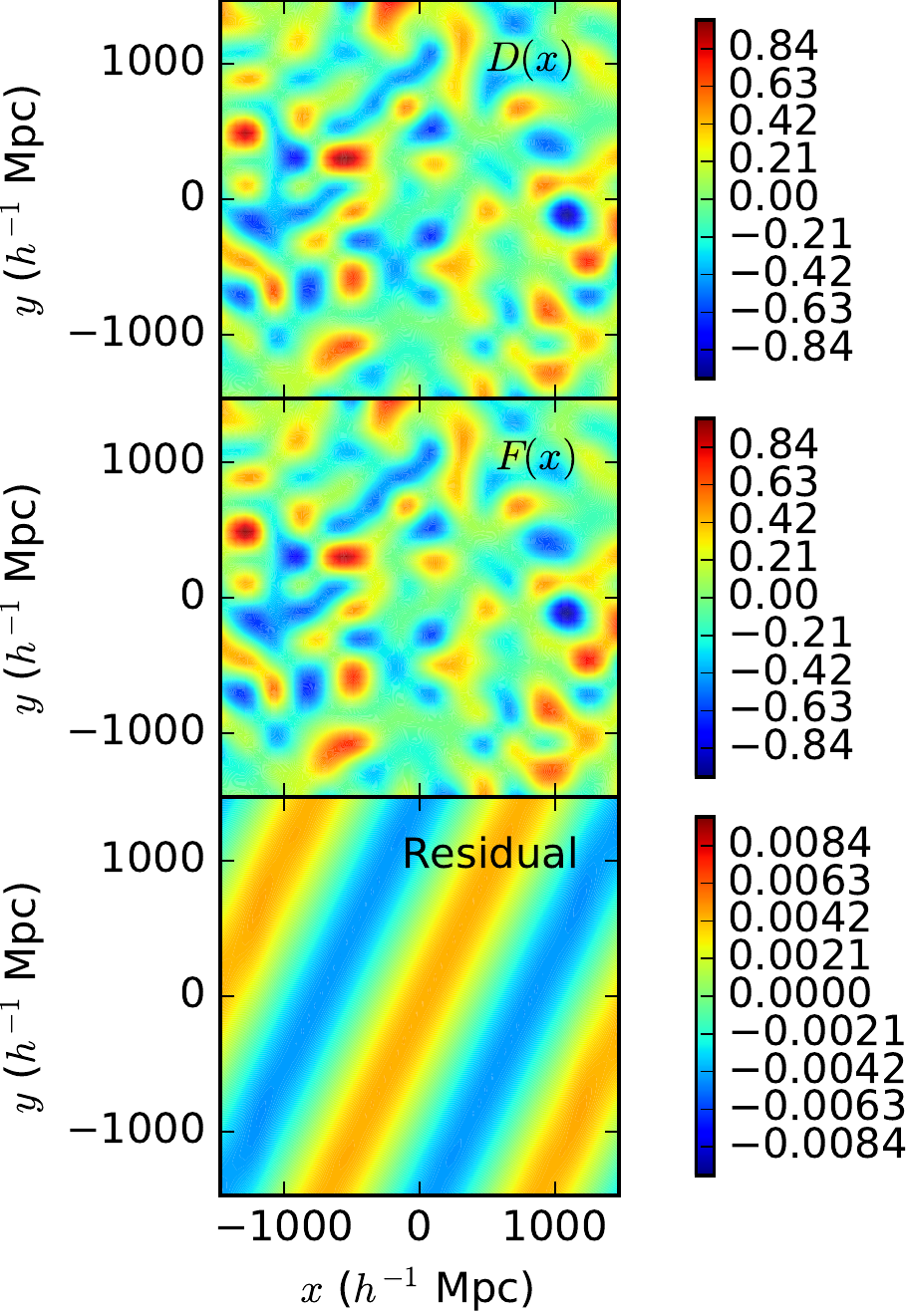}
\caption{This plot is similar to Fig.~\ref{fig:spikefield}, but shows a slice through a field with a single contaminated mode as described in Sec.~\ref{sec:scm}. The best-fitting $\epsilon^\mathrm{(BF)}$ for this particular realisation amounts to 1.005. All panels appear very similar; this is quantified in Fig.~\ref{fig:highlybiased}.}
\label{fig:ex2field}
\end{figure}

As a second example we use Eq.~\eqref{eq:debiasfactor} to construct a contaminant that would lead to a strong bias in the recovered $P(k)$ without the debiasing step. Eq.~\eqref{eq:debiasfactor} only contains positive quantities and is normalised such that the bias is a value between 0 and 1. 1 corresponds to an unbiased estimate, hence 0 is the maximal bias. This extreme case would be fulfilled if $f$ is large for one mode and 0 otherwise. Therefore, we construct a contaminant that is a large number at the modes corresponding to $\k=\pm (0.003,0.003,0.003)h\;\mathrm{Mpc}^{-1}$. An example of this setup can be found in Fig.~\ref{fig:ex2field}. The top panel again shows an uncontaminated Gaussian random field, the central panel shows the same field with the contaminant added. The contaminant itself is not as prominent as the one in Fig.~\ref{fig:spikefield}, because this single contaminated mode just adds a long wavelength contribution in real space. The bottom panel shows the field after subtracting the template.

\begin{figure}
\centering
\includegraphics[width=\linewidth]{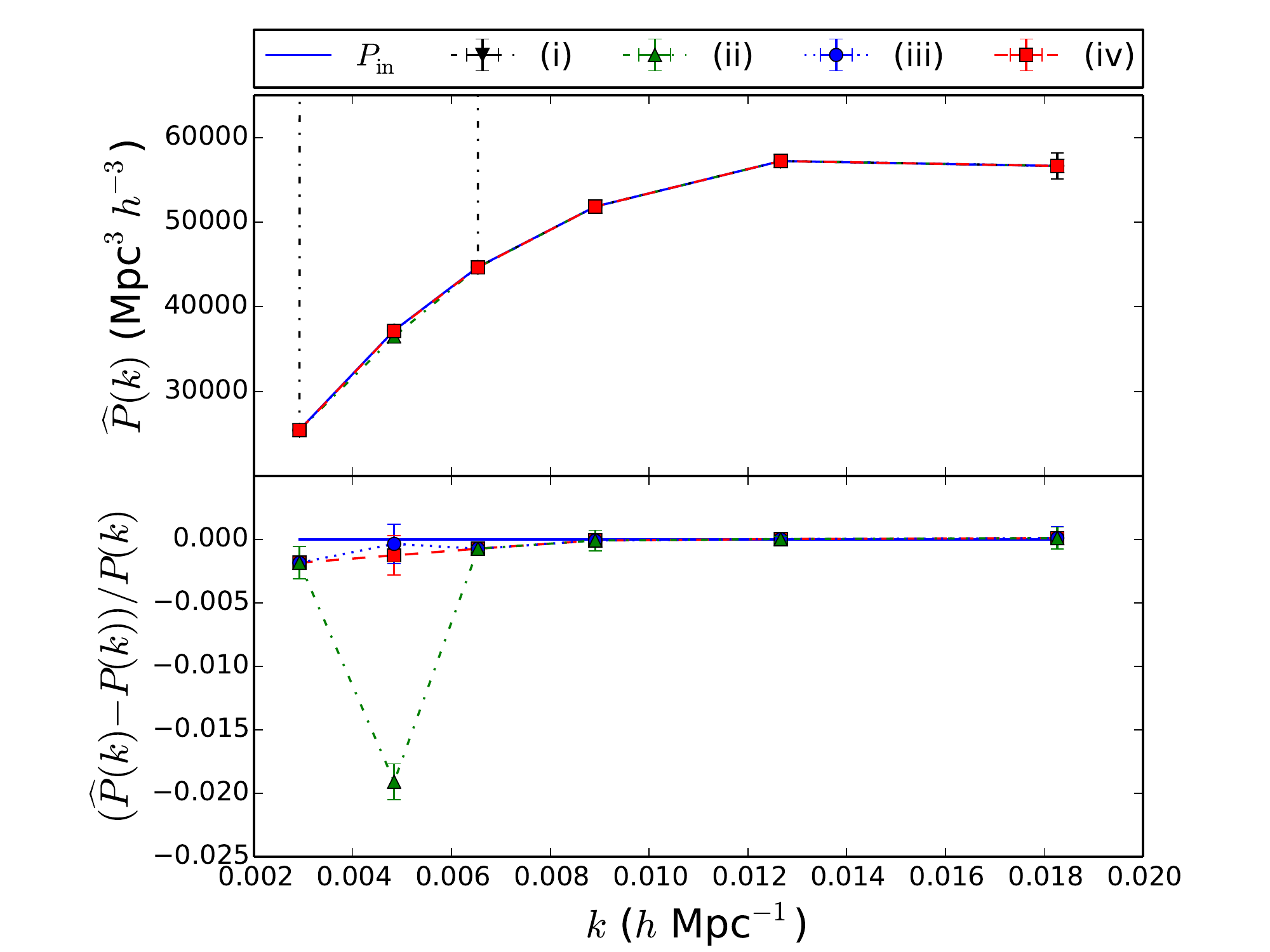}
\caption{Means and standard deviations of the power spectra of 1000 realisations of Gaussian random fields contaminated with Hermitian Gaussian spikes. The red dots represent measurements, where the contamination has not been taken into account. For the blue dots, \textit{mode deprojection} has been used to remove the spikes. For the green dots, we used \textit{debiased \revision{mode} subtraction}. The solid blue line shows the input power spectrum.}
\label{fig:highlybiased}
\end{figure}

We measure the same cases (i)-(iv) as in the previous subsection, which we plot in Fig.~\ref{fig:highlybiased}. The prior power is again the input power. If we were to apply this to a real survey, we would not know the true power, so we perform a few runs, where we first assume a flat prior power spectrum $P(k)=1\;\forall k$, and then iteratively compute the power with the power from the previous run as the prior power spectrum. The effect of the prior power spectrum is negligible, because the result in the first step provides the same result as assuming the input power as prior.

The data points for all cases (i)-(iv) are close to the input power in all bins but the second. In the second bin, the power spectrum for case (i) extends beyond the plotted range, chosen to highlight differences between the other approaches. In case (ii), the power is significantly underestimated. The bias amounts to about 2 per cent, i.e. it highly affects measurements where small-$\k$ modes are crucial, such as $f_\mathrm{NL}$-measurements. The difference between the cases (iii) and (iv) is much smaller, even in this extreme example.

\section{Conclusions}
\label{sec:conclusion}
We have considered methods to remove contaminants when measuring the 3D galaxy power spectrum from a given density field, focussing on \textit{mode deprojection} and \revision{\textit{mode subtraction}}. In order to understand how these are related, we have decomposed the problem into separate steps. In particular we have separated \textit{mode deprojection} from power spectrum estimation - they are often considered together - arguing that this split makes sense given the mathematical equivalence of \textit{mode deprojection} and \revision{\textit{mode subtraction}}. We argue that the QML estimation is not practical for modern surveys with large numbers of observed modes, but that we can apply \textit{mode deprojection} to the FKP-estimator, using the mathematical equivalence of \textit{mode deprojection} and \revision{\textit{mode subtraction}}, thus avoiding having to create large estimator and covariance matrices for all modes. The resulting estimate is biased, but can easily be made unbiased with a simple correction, again that can be implemented without the \revision{inversion} of large matrices. This correction is easily extended to the case of multiple contaminants and is not affected if the modes are correlated even without the effects of contaminants. The final result of our short paper is the suggestion that 3D galaxy power spectrum should be estimated using Eq.~\eqref{eq:ubPTS},

\begin{equation}
	\widehat{P}(k_i)=\frac{1}{N_{\Bbbk_i}}\sum_{\k_\alpha}\frac{\left\vert F(\k_\alpha)-\frac{S_P}{R_P}f(\k_\alpha)\right\vert^2}{1-\frac{1}{R_P}\frac{\vert f(\k_\alpha)\vert^2}{P(k_\alpha)}}.
\end{equation}
While theoretically it is sub-optimal, in practice the degradation of signal is expected to be less than ignoring window effects in the optimisation of mode averaging when using the standard FKP estimator.

\section*{Acknowledgments}

The authors would like to thank Franz Elsner, H\'ector Gil-Mar\'in, Ashley Ross and the unknown referee for valuable comments.

We used matplotlib \citep{Hunter:2007} to generate plots. The CAMB package \citep{Lewis:1999bs} has been used to generate model and prior power spectra. We made use of the facilities and staff
of the UK Sciama High Performance Computing cluster supported
by the ICG, SEPNet and the University of Portsmouth.

BK thanks the Faculty of Technology of the University of Portsmouth for support during his PhD studies. WJP and DB acknowledge support from UK STFC through the consolidated grant ST/K0090X/1, WJP also acknowledges support from the European Research
Council through the Darksurvey grant and the UK Space Agency through grant ST/N00180X/1. LS is grateful for support from SNSF grant SCOPES IZ73Z0-152581, GNSF grant FR/339/6-350/14,
and DOE grant DEFG 03-99EP41093. 

\bibliographystyle{mnras}
\bibliography{systpap2}

\appendix

\section{Derivation of Mode Deprojection with multiple templates}
\label{sec:MPmultderiv}

In this appendix we want to derive Eq.~\eqref{eq:multiBMPpower} from Eq.~\eqref{eq:MPmult}. We start by rewriting Eq.~\eqref{eq:MPmult} in matrix notation
\begin{equation}
	\tilde\C=\C+\lim_{\sigma\rightarrow\infty}\sigma\f\mathbf{I}_{N_\mathrm{sys}}\f^\dagger, 
\end{equation}
defining an $N_\mathrm{mode}\times N_\mathrm{sys}$ matrix $\f_{\alpha A}\equiv f_A(\k_\alpha)$, such that we can invert $\tilde\C$ using the Woodbury matrix identity
\begin{align}
	\tilde\C\inv&=\C\inv-\C\inv\lim_{\sigma\rightarrow\infty}\sigma\f\left(\mathbf{I}\inv_{N_\mathrm{sys}}+\f^\dagger\C\inv\sigma\f\right)\inv\f^\dagger\C\inv\nonumber\\
	&=\C\inv-\C\inv\f\left(\f^\dagger\C\inv\f\right)\inv\f^\dagger\C\inv\nonumber\\
	&\equiv\C\inv-\C\inv\f\mathbf{R}\inv\f^\dagger\C\inv.
\end{align}
If we assume $\C_{\alpha\beta}=\delta_{\alpha\beta}P(k_\alpha)$, $\mathbf{R}\equiv\f^\dagger\C\inv\f$ becomes a matrix equivalent to the factor $R_P$ in previous sections:
\begin{equation}
	\mathbf{R}_{AB}=\sum_{\mu\nu}f_A^\ast(\k_\mu)\frac{\delta_{\mu\nu}}{P(k_\mu)}f_B(k_\nu)=\sum_{\mu}\frac{f_A^\ast(\k_\mu)f_B(k_\mu)}{P(k_\mu)}.
\end{equation}
The inverse updated covariance matrix then reads
\begin{equation}
	\tilde\C\inv_{\alpha\beta}=\frac{\delta_{\alpha\beta}}{P(k_\alpha)}-\sum_{AB}\frac{f_A(\k_\alpha)\mathbf{R}\inv_{AB}f_B^\ast(\k_\beta)}{P(k_\alpha)P(k_\beta)}.
	\label{eq:CinvtildeMulti}
\end{equation}
If we do not bin, but apply mode deprojection to each mode separately, the matrix $\tilde\E$ simplifies to
\begin{equation}
	\tilde\E_{\alpha\beta}(k_j)=\sum_{\mu\nu}\tilde\C\inv_{\alpha\mu}\delta_{\mu j}\delta_{\mu\nu}\tilde\C\inv_{\nu\beta}=\tilde\C\inv_{\alpha j}\tilde\C\inv_{j\beta}.
	\label{eq:Etildemulti}
\end{equation}
After inserting Eq.~\eqref{eq:CinvtildeMulti} into Eq.~\eqref{eq:Etildemulti}, we obtain
\begin{align}
	P^2&(k_j)\sum_{\alpha\beta} F^\ast(\k_\alpha)\tilde\E_{\alpha\beta}(k_j)F(\k_\beta)\nonumber\\
	=&\vert F(\k_j)\vert^2
	\nonumber\\
	&-\sum_{AB\alpha}\frac{F^\ast(\k_\alpha)f_A(\k_\alpha)}{P(k_\alpha)}\mathbf{R}\inv_{AB}f_B^\ast(\k_j)F(\k_j)\nonumber\\
	&-\sum_{AB\beta}F^\ast(\k_j)f_A(\k_j)\mathbf{R}\inv_{AB}\frac{f_B^\ast(\k_\beta)F(\k_\beta)}{P(k_\beta)}\nonumber\\
	&+\sum_{ABCD\alpha\beta}\frac{F^\ast(\k_\alpha)f_A(\k_\alpha)}{P(k_\alpha)}\mathrm{R}\inv_{AB}f^\ast_B(\k_j)f_C(\k_j)\mathrm{R}\inv_{CD}\frac{f_D^\ast(\k_\beta)F(\k_\beta)}{P(k_\beta)}\nonumber\\
	=&\vert F(\k_j)\vert^2
	\nonumber\\
	&-2\Re\left[\sum_{AB}\mathbf{S}_A\mathbf{R}\inv_{AB}f_B^\ast(\k_j)F(\k_j)\right]\nonumber\\
	&+\left\vert\sum_{AB}\mathbf{S}_A\mathbf{R}\inv_{AB}f_B(\k_j)\right\vert^2\nonumber\\
	=&\left\vert F(\k_j)-\sum_{AB}\mathbf{S}_A\mathbf{R}\inv_{AB}f_B(\k_j)\right\vert^2,
\end{align}
where we defined $\mathbf{S}_A\equiv \sum_\alpha\frac{f_A(\k_\alpha)F^\ast(\k_\alpha)}{P(k_\alpha)}$ analogous to $S_P$.

\section{Derivation of \revision{mode} subtraction with multiple templates}
\label{sec:TSmultideriv}
Here we derive the best-fitting $\bvarepsilon^{(BF)}$ from the likelihood
\begin{equation}
	-2\ln\mathcal{L}=\sum_\alpha\frac{\vert F(\k_\alpha)-\sum_A \mathbf{\varepsilon}_A f_A(\k_\alpha)\vert^2}{P(k_\alpha)}
\end{equation}
to find the same result as in the previous appendix.
Taking the derivative with respect to $\bvarepsilon_B$ yields
\begin{equation}
	\frac{\partial\chi^2}{\partial\bvarepsilon_B}=-2\bvarepsilon_B\sum_\alpha\frac{f_B(\k_\alpha)F^\ast(\k_\alpha)-\sum_A\bvarepsilon_A f_B(\k_\alpha)f^\ast_A(\k_\alpha)}{P(k_\alpha)}.
\end{equation}
This derivative is zero if 
\begin{equation}
	\sum_\alpha\frac{f_B(\k_\alpha)F^\ast(\k_\alpha)}{P(k_\alpha)}=\sum_{A\alpha}\frac{\bvarepsilon_A f_B(\k_\alpha)f^\ast_A(\k_\alpha)}{P(k_\alpha)},
\end{equation}
which reads
\begin{equation}
	\mathbf{S}=\mathbf{R}\bvarepsilon
\end{equation}
in matrix notation. The best fitting $\bvarepsilon$ value is therefore given by
\begin{equation}
	\bvarepsilon^\mathrm{(BF)}=\mathbf{R}\inv\mathbf{S}.
\end{equation}
The absolute value squared of the best fitting signal is hence equal to Eq.~\eqref{eq:multiBMPpower}:
\begin{align}
	\left\vert F(\k_\alpha)-\sum_{A}\bvarepsilon^\mathrm{(BF)}_A f_A(\k_\alpha)\right\vert^2=\left\vert F(\k_\alpha)-\sum_{AB}\mathbf{R}\inv_{AB}\mathbf{S}_B f_A(\k_\alpha)\right\vert^2.
\end{align}

\section{\revision{mode} subtraction and the Debiasing Step for a non-diagonal Covariance Matrix}

\revision{\subsection{Including the covariance in the calculation of the best-fit mode to subtract}
\label{sec:nondiag}}
Here we show a derivation similar to the one in Sec.~\ref{sec:systrem} and \ref{sec:debias} for the more general case of a non-diagonal covariance matrix. We show that the debiasing works in the same way as in Sec.~\ref{sec:debias}, just with a generalised definition of $R_P$.

Defining the covariance matrix of the true density
\begin{equation}
	\C_{\alpha\beta}\equiv\left\langle D_\alpha D_\beta^\ast\right\rangle
\end{equation}
and assuming that the true signal and the contaminant are uncorrelated, we can write
\begin{equation}
	\left\langle F_\alpha F_\beta^\ast\right\rangle=\C_{\alpha\beta}+\revision{\varepsilon_\mathrm{true}^2} f_\alpha f_\beta^\ast.
	\label{eq:Cdef}
\end{equation}
As we did in Sec.~\ref{sec:systrem}, we introduce a free parameter $\varepsilon$, such that 
\begin{equation}
	\revision{\widehat D}_\alpha\equiv F_\alpha-\varepsilon f_\alpha.
\end{equation}
Assuming that the true density field is Gaussian, its log-likelihood reads
\begin{equation}
	-2\ln\mathcal{L}=\sum_{\alpha\beta}\left(F_\alpha-\varepsilon f_\alpha\right)^\ast \C_{\alpha\beta}\inv\left(F_\beta-\varepsilon f_\beta\right)+\mathrm{const}.
\end{equation}
To find the best-fitting $\varepsilon^\mathrm{(BF)}$, we take the derivative of the log-likelihood with respect to $\varepsilon$:
\begin{align}
	-2\frac{\partial\ln\mathcal{L}}{\partial \varepsilon}=&-\sum_{\alpha\beta}f_\alpha^\ast \C_{\alpha\beta}\inv\left(F_\beta-\varepsilon f_\beta\right)\nonumber\\
	&-\sum_{\alpha\beta}\left(F_\alpha-\varepsilon f_\alpha\right)^\ast \C_{\alpha\beta}\inv f_\beta\nonumber\\
	=&2\varepsilon\sum_{\alpha\beta}f_\alpha^\ast \C_{\alpha\beta}\inv f_\beta\nonumber\\
	&-\sum_{\alpha\beta}\left[f_\alpha^\ast \C_{\alpha\beta}\inv F_\beta+F_\alpha^\ast \C_{\alpha\beta}\inv f_\beta\right].
	\label{eq:logLderiv}
\end{align}
As $\C$ is a covariance matrix of complex random variables, it is Hermitian positive-semidefinite, such that the second sum can be written as
\begin{equation}
	\sum_{\alpha\beta}\left[f_\alpha^\ast \C_{\alpha\beta}\inv F_\beta+F_\alpha^\ast \C_{\alpha\beta}\inv f_\beta\right]=2\Re\left[\sum_{\alpha\beta}f_\alpha^\ast \C_{\alpha\beta}\inv F_\beta\right].
\end{equation}
For shortness and in analogy to Sec.~\ref{sec:systrem}, we call this sum 
\begin{equation}
	S_P\equiv \sum_{\alpha\beta}\Re\left[f_\alpha^\ast \C_{\alpha\beta}\inv F_\beta\right]
\end{equation}
and the first sum in Eq.~\eqref{eq:logLderiv} we call
\begin{equation}
	R_P\equiv \sum_{\alpha\beta}f_\alpha^\ast \C_{\alpha\beta}\inv f_\beta.
	\label{eq:RPdef}
\end{equation}
We obtain the best-fitting, i.e. maximum likelihood, value 
\begin{equation}
	\varepsilon^\mathrm{(BF)}=\frac{S_P}{R_P}
\end{equation}
by equating Eq.~\eqref{eq:logLderiv} to zero.

Now we want to calculate the expectation value 
\begin{align}
	\left\langle\left\vert F_\alpha-\frac{S_P}{R_P} f_\alpha\right\vert^2\right\rangle=&\left\langle\left\vert F_\alpha\right\vert^2\right\rangle-\frac{2}{R_P}\Re\left[\left\langle S_P F_\alpha^\ast f_\alpha\right\rangle\right]\nonumber\\
	&+\left\langle\frac{S_P^2}{R_P^2}\right\rangle\left\vert f_\alpha\right\vert^2.
	\label{eq:ExpVal}
\end{align}
We calculate each term separately: 
\begin{enumerate}
	\item The first term $\langle\vert F_\alpha\vert^2\rangle=\C_{\alpha\alpha}+\revision{\varepsilon_\mathrm{true}^2}\vert f_\alpha\vert^2$ is a special case of Eq.~\eqref{eq:Cdef}.
	\item To calculate the second term, we reexpand $S_P$ and use the fact that $\Re\left[F_\alpha f_\alpha^\ast\right]=\Re\left[F_\alpha^\ast f_\alpha\right]$: \begin{align} 2\Re\left[\left\langle S_P F_\alpha^\ast f_\alpha\right\rangle\right]=2\Re\left[\sum_{\gamma\beta}f_\gamma^\ast \C_{\gamma\beta}\inv \left\langle F_\beta F_\alpha^\ast\right\rangle f_\alpha\right] \end{align}
	After reinserting Eq.~\eqref{eq:Cdef}, we get
	\begin{align}
		2\Re\left[\left\langle S_P F_\alpha^\ast f_\alpha\right\rangle\right]=&2\Re\left[\sum_{\gamma\beta}f_\gamma^\ast \C_{\gamma\beta}\inv \C_{\beta\alpha} f_\alpha\right.\nonumber\\
		&\left.+\revision{\varepsilon_\mathrm{true}^2}\sum_{\gamma\beta}f_\gamma^\ast \C_{\gamma\beta}\inv f_\beta f_\alpha^\ast f_\alpha\right].
	\end{align}
	In the first term we have $\sum_{\beta}\C_{\gamma\beta}\inv \C_{\beta\alpha}=\delta_{\gamma\alpha}$, and in the second term we find the definition of $R_P$. Thus, the second term of Eq.~\eqref{eq:ExpVal} is
	\begin{equation}
		2\Re\left[\left\langle S_P F_\alpha^\ast f_\alpha\right\rangle\right]=2\vert f_\alpha\vert^2 \left(1+\revision{\varepsilon_\mathrm{true}^2}R_P\right).
	\end{equation}
	\item In the third term, we can again make use of Eq.~\eqref{eq:Cdef}:
		\begin{align}
			\left\langle S_P^2\right\rangle&=\sum_{\alpha\beta\gamma\delta} \Re\left[f_\alpha^\ast\C_{\alpha\beta}\inv f_\gamma \C_{\gamma\delta}\inv\left\langle F_\beta F_\delta^\ast\right\rangle\right]\nonumber\\
			&=\sum_{\alpha\beta\gamma\delta} \Re\left[f_\alpha^\ast\C_{\alpha\beta}\inv f_\gamma \C_{\gamma\delta}\inv\C_{\beta\delta}+\revision{\varepsilon_\mathrm{true}^2}f_\alpha^\ast\C_{\alpha\beta}\inv f_\gamma \C_{\gamma\delta}\inv f_\beta f_\delta^\ast\right]
		\end{align}
		In the first term, we have again $\sum_{\beta}\C_{\alpha\beta}\inv \C_{\beta\delta}=\delta_{\alpha\delta}$, and the second term is equal to $R_P^2$, such that
		\begin{align}
			\left\langle S_P^2\right\rangle&=\sum_{\alpha\gamma} \Re\left[f_\alpha^\ast f_\gamma \C_{\gamma\alpha}\inv\right]+\revision{\varepsilon_\mathrm{true}^2}R_P^2=R_P+\revision{\varepsilon_\mathrm{true}^2}R_P^2
		\end{align}
\end{enumerate}
Recollecting 1.-3. and inserting into Eq.~\eqref{eq:ExpVal} yields
\begin{align}
	\left\langle\left\vert F_\alpha-\frac{S_P}{R_P} f_\alpha\right\vert^2\right\rangle=&\C_{\alpha\alpha}+\revision{\varepsilon_\mathrm{true}^2}\vert f_\alpha\vert^2\nonumber\\
	&-2\vert f_\alpha\vert^2 \left(\revision{\varepsilon_\mathrm{true}^2}+\frac{1}{R_P}\right)+\left(\revision{\varepsilon_\mathrm{true}^2}+\frac{1}{R_P}\right)\left\vert f_\alpha\right\vert^2\nonumber\\
	&=\C_{\alpha\alpha}-\frac{\vert f_\alpha\vert^2}{R_P}.
\end{align}
As the power spectrum $P(k_\alpha)=\C_{\alpha\alpha}$ is defined as the diagonal elements of the covariance matrix, the debiasing step is the same for a non-diagonal covariance matrix as for a diagonal one (cf. Sec.~\ref{sec:debias}), we just have to use the generalised definition of $R_P$ as in Eq.~\eqref{eq:RPdef}.

\revision{\subsection{The independent mode approximation}
\label{sec:diagCapprox}
We have seen in Appendix \ref{sec:nondiag} that \revision{mode} subtraction also works when the covariance matrix is non-diagonal. However, to compute the generalised $R_P$, one has to invert the full $N_\mathrm{mode}\times N_\mathrm{mode}$ covariance matrix, which makes this approach computationally almost as expensive as using the QML estimator. We will argue that, in most cases, Eq.~\eqref{eq:ubPTS} provides a good estimate of the power, even in the presence of covariant modes, and we will provide a further correction term that corrects for using Eq.~\eqref{eq:ubPTS} when off-diagonal covariances are important.

Suppose we apply Eq.~\eqref{eq:ubPTS} assuming a diagonal covariance matrix, even though there are covariances between different modes. Then, we find a best fitting 
\begin{equation}
	\varepsilon_\mathrm{BF}^\prime=\frac{\sum_\alpha\frac{F^\ast_\alpha f_\alpha}{P_\alpha}}
	{\sum_\mu\frac{\vert f_\mu\vert^2}{P_\mu}}
\end{equation}
instead of the true 
\begin{equation}
	\varepsilon_\mathrm{BF}=\frac{\sum_{\alpha\beta}f_\alpha^\ast \C_{\alpha\beta}\inv F_\beta}{\sum_{\alpha\beta}f_\alpha^\ast \C_{\alpha\beta}\inv f_\beta}.
\end{equation}
The expectations are the same $\left\langle\varepsilon_\mathrm{BF}^\prime\right\rangle=\left\langle\varepsilon_\mathrm{BF}\right\rangle=\varepsilon_\mathrm{true}$, but their variances are different. For the approximate estimate we have
\begin{align}
	\left\langle\varepsilon_\mathrm{BF}^{\prime 2}\right\rangle=&\frac{\left\langle \sum_{\alpha\beta}\frac{F^\ast_\alpha f_\alpha F_\beta f^\ast_\beta}{P_\alpha P_\beta}\right\rangle}{R_P^{\prime 2}}\nonumber\\
	=&\frac{\sum_{\alpha\beta}\frac{ f_\alpha\C_{\alpha\beta} f^\ast_\beta}{P_\alpha P_\beta}}{R_P^{\prime 2}}+\frac{\sum_{\alpha\beta}\frac{\varepsilon_\mathrm{true}^2 \left\vert f_\alpha\right\vert^2 \left\vert f_\beta\right\vert^2}{P_\alpha P_\beta}}{R_P^{\prime 2}}\nonumber\\
	=&\frac{1}{R_P^{\prime 2}}\sum_{\alpha\beta}\frac{ f_\alpha\C_{\alpha\beta} f^\ast_\beta}{P_\alpha P_\beta}+\varepsilon_\mathrm{true}^2.
\end{align}
Unlike in the previous estimates, the covariance matrix does not cancel in the first term. Similarly, 
\begin{align}
	\left\langle\varepsilon_\mathrm{BF}^\prime F_\alpha^\ast f_\alpha\right\rangle=&\frac{1}{R_P^\prime}\left\langle\sum_\beta\frac{f_\beta^\ast F_\beta F_\alpha^\ast f_\alpha}{P_\beta}\right\rangle\nonumber\\
	=&\frac{1}{R_P^{\prime}}\sum_{\beta}\frac{ f_\alpha\C_{\alpha\beta} f^\ast_\beta}{P_\beta}+\varepsilon_\mathrm{true}^2\left\vert f_\alpha\right\vert^2.
\end{align} 
Combining the previous two equations, we obtain
\begin{align}
	\left\langle\left\vert F_\alpha-\varepsilon_\mathrm{BF}^\prime f_\alpha\right\vert^2\right\rangle=C_{\alpha\alpha}-\frac{2}{R_P^\prime}\sum_{\beta}\frac{ f_\alpha\C_{\alpha\beta} f^\ast_\beta}{P_\beta}+\frac{\vert f_\alpha\vert^2}{R_P^{\prime 2}}\sum_{\gamma\beta}\frac{ f_\gamma\C_{\gamma\beta} f^\ast_\beta}{P_\gamma P_\beta}.
\end{align}
Splitting the covariance matrix 
\begin{equation}
	C_{\alpha\beta}=P_\beta\left(\delta_{\alpha\beta}+\Delta_{\alpha\beta}\right)
\end{equation}
into a diagonal and off-diagonal elements yields
\begin{align}
	\left\langle\left\vert F_\alpha-\varepsilon_\mathrm{BF}^\prime f_\alpha\right\vert^2\right\rangle=P_\alpha-\frac{\vert f_\alpha\vert^2}{R_P^\prime}\left[1+\sum_{\gamma\beta} f_\gamma\Delta_{\gamma\beta} f^\ast_\beta\left(\frac{2\delta_{\alpha\gamma}}{\vert f_\alpha\vert^2}
	-\frac{1}{R_P^{\prime}P_\gamma}\right)\right].
\end{align}
Hence, one can perform \revision{mode} subtraction assuming a diagonal covariance matrix and then apply another correction term which is linear in its off-diagonal elements. The advantage of this procedure is that it does not require any inversion of the $N_\mathrm{mode}^2$ covariance matrix. If the off-diagonal elements are small, then the bias correction reverts back to the form of Eq. \eqref{eq:debiasfactor}.
}

\bsp

\label{lastpage}

\end{document}